\documentclass[12pt]{article}

\usepackage{graphicx}

\large
\newcommand{\be}{\begin{eqnarray}}
\newcommand{\ee}{\end{eqnarray}}

\newcommand{\ba}{\begin{array}}
\newcommand{\ea}{\end{array}}

\newcommand{\el}{\nonumber \hfill \\}

\begin{document}
\setlength{\baselineskip}{21pt}
\pagestyle{empty}
\vfill

\vskip 1.0cm
\centerline{\Large
\bf  Robustness and Optimality of Light Harvesting}
\centerline{\Large
\bf in Cyanobacterial Photosystem~I}
\vskip 0.5cm
\centerline{Melih K. \c Sener ${}^{1}$,
Deyu Lu ${}^{1,2}$,
Thorsten Ritz ${}^{3}$,
Sanghyun Park ${}^{1,2}$,
Petra Fromme ${}^{4,5}$,
and Klaus Schulten ${}^{1,2,*}$}
\vskip 0.2cm
\noindent
{\small
${}^{1}$ Beckman Institute, University of Illinois
at Urbana-Champaign,
Urbana, Illinois 61801 \\
 ${}^{2}$ Department of Physics, University of Illinois
 at Urbana-Champaign,
Urbana, Illinois 61801 \\
${}^{3}$  Department of Biology,
Virginia Polytechnic Institute and State University,
Blacksburg, \hbox{VA 24061}  \\
${}^{4}$ Max-Volmer Laboratorium f\" ur Biophysikalische
Chemie, Institut f\"{u}r Chemie, Fakult\"{a}t 2; \\
Technische Universit\" at Berlin,
D-10623 Berlin Germany  \\
 ${}^{5}$ Department of Chemistry and Biochemistry, Arizona State University, Tempe,
AZ 85287-1604 \\
 ${}^{*}$ corresponding author: kschulte@ks.uiuc.edu
}
\vskip 0.3cm

\centerline{(July 15, 2002)}


\centerline{(Accepted for publication in {\em J.\ Phys.\ Chem.\ B} 2002.)}

\vskip 0.3cm

\centerline{\bf Abstract}

As most biological species, photosynthetic
lifeforms have evolved to function optimally,
despite thermal disorder
and with fault tolerance.
It remains a challenge to understand
how this is achieved.
To address this challenge
the function of
the protein-pigment complex photosystem~I (PSI)
of the cyanobacterium {\em Synechococcus elongatus}
is investigated theoretically.
The recently obtained high resolution structure
of this complex exhibits an aggregate of 96 chlorophylls
that are electronically coupled to function as a
light-harvesting antenna complex.
This paper constructs an
effective Hamiltonian for
the chlorophyll aggregate
to describe
excitation transfer
dynamics and spectral properties of PSI.
For this purpose,
a new kinetic expansion
method, the sojourn expansion, 
is introduced.
Our study
shows that at room temperature fluctuations
of site energies have little effect
on the calculated excitation lifetime and quantum yield,
which compare favorably with experimental
results.
The efficiency of the system is found to be robust
against `pruning' of individual chlorophylls.
An optimality
of the arrangement of chlorophylls
is identified through
the quantum yield in
comparison with
an ensemble of randomly
oriented chlorophylls,
though, 
the quantum yield is seen to change only within a
narrow interval in such an ensemble.

\vfill
\noindent

\eject
\pagestyle{plain}

\noindent
\section{Introduction}


Photosynthesis is the main source of energy in our
biosphere. Although light-harvesting systems show
a great variety in their design and function,
most of them are membrane proteins comprised of a network of
pigment antennae surrounding
a reaction center~\cite{BLAN2002}.
The antenna complex is responsible for the absorption
of light and the transfer of the resulting electronic
excitation to
a so-called special pair of chlorophylls in
the reaction center, where a charge separation
across the cell membrane is initiated.
This charge separation is later utilized by the cell to store energy
through synthesis of ATP
in a more stable form.

Oxygenic photosynthetic species employ two transmembrane
protein-pigment complexes for light-harvesting, named
photosystems I and II.
Photosystem I (PSI) is a ubiquitous protein-pigment complex found
in green plants, algae and cyanobacteria.
Located in the bacterial membrane, it absorbs
sun light and uses its
energy to transfer electrons across the
cell membrane.
PSI is remarkable in that it contains the antenna complex,
reaction center and the electron transfer chain
all within the same protein.
For a recent review on PSI physiology and function
we refer the reader to Refs.~\cite{GOBE2001,CHIT2001,MELK2002}.

A high resolution atomic structure
was recently obtained for PSI from the
cyanobacterium
{\em Synechococcus (S.) elongatus}~\cite{JORD2001}.
The structure exhibits an aggregate
of 96 chlorophyll
molecules, including the special pair,
P700, at which the primary charge
separation takes place.
This finding opens up the opportunity to
study in detail
the light-harvesting function of PSI,
i.e., one
can computationally examine the light-harvesting
pathway from
initial absorption of a photon to the
transfer of electrons by the special pair.
In the past, a similar leap in structural information on
the peripheral light-harvesting complexes
LH1 and LH2 of purple bacteria resulted
in the photosynthetic unit of purple bacteria
being the best understood
antenna complex
today~\cite{MCDE95,MILL82,KOEP96,SUND99,HU97}.

\begin{figure}
\centerline{\includegraphics[width=8.25cm]{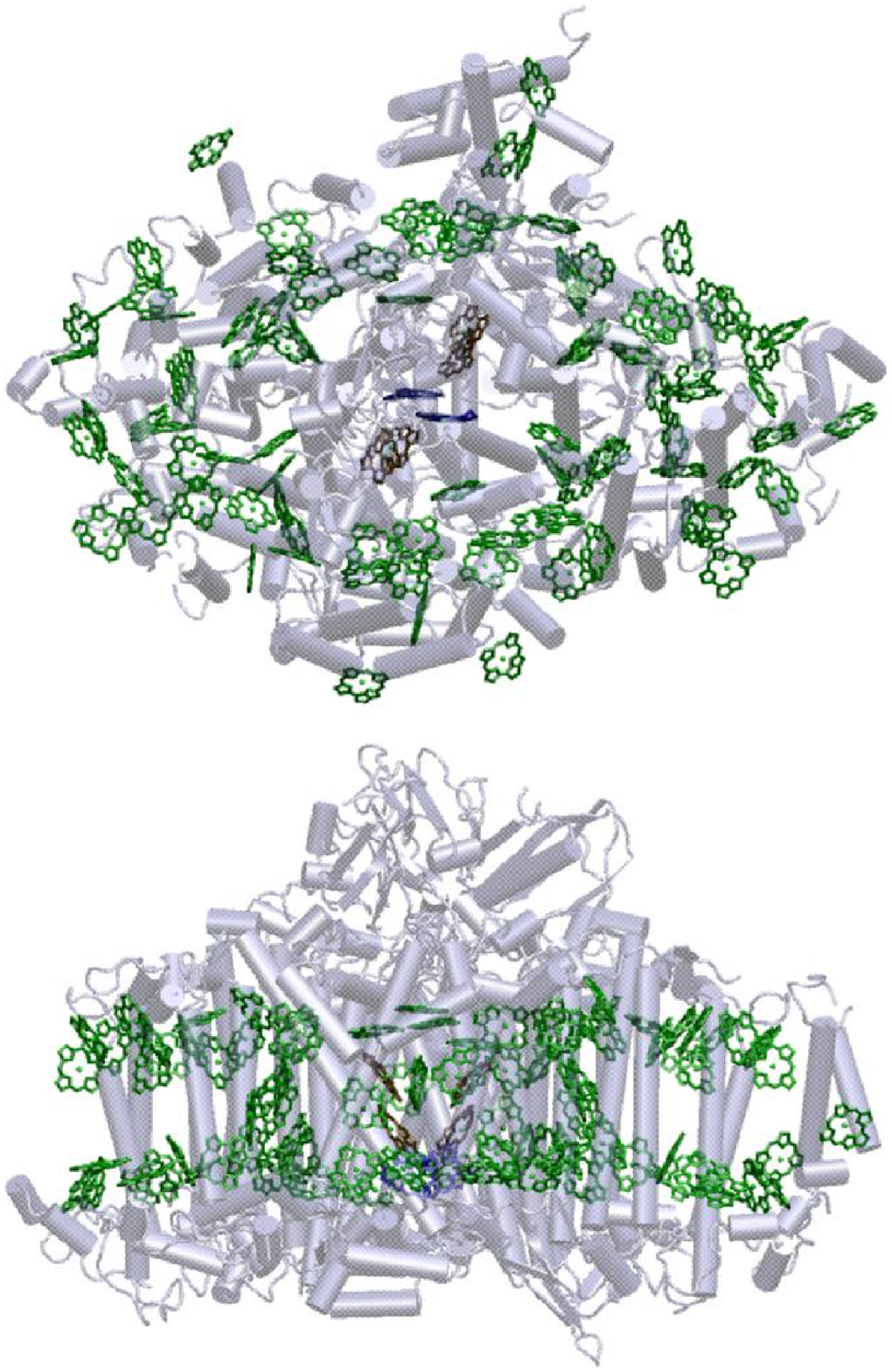}}
\caption[ps1fig]{Photosystem I of the cyanobacterium
{\em Synechococcus elongatus} and its 96 chlorophylls.
The special pair P700 is shown in blue. The remaining four
reaction center chlorophylls are shown in red.
For clarity, neither
the chlorophyll tails nor other cofactors
are shown.
The chlorophyll pool surrounding the reaction
center displays a pseudo-$C_2$ symmetry.
Figure produced with VMD~\cite{HUMP96}.}
\label{fig:ps1_chl_prot}
\end{figure}

It is of interest to compare the structure and function
of PSI to the photosynthetic unit of purple
bacteria. Both systems contain on the order of a hundred
chromophores per reaction center.
In purple bacteria these chromophores are
organized in highly symmetric ring-like
structures~\cite{HU2002,RITZ2002,SCHU99}.
In contrast,
the chlorophyll aggregate in PSI forms a
rather random
array surrounding the reaction center
(see Fig.~\ref{fig:ps1_chl_prot}),
the only hint of order being
a pseudo-$C_2$ symmetry apparent in the structure.

Purple bacteria have the oldest
known photosynthetic apparatus, while
cyanobacteria and their relatives
are evolutionarily
more recent~\cite{XION2000,BLAN2001}.
It is
of interest to compare the light-harvesting systems
of these lifeforms and to elucidate
the drive toward a more efficient system in
the more recently developed light-harvesting complexes.

A related issue is whether the peculiar
arrangement of chlorophylls
found in the reported structure
is essential for
the functioning of PSI or whether
the system can be described
successfully
assuming some degree of randomness.
For example,
is every single chlorophyll essential for the
proper functioning of the system?
This question also constitutes
a modeling challenge.
Due to its circular symmetry
and repetitive elements
a purple bacterial light-harvesting complex
can be modeled
with only a few
physical parameters
describing
chlorophyll
site energies and
electronic couplings~\cite{HU97,RITZ98,DAMJ2000}.
PSI, on the other hand, features 96 chlorophylls
that show few regularities
and seemingly require a large number of parameters
for their description.

There are two possible ways
to approach the complexity
of the chlorophyll aggregate in PSI.
In one approach,
typically taken in chemistry,
one strives for
an ever increasing accuracy in a model through
a large number of specific parameters.
This approach implies that
the specific properties of every one
of the 96 chlorophylls in PSI
matter.
In another approach,
typically taken in physics,
one identifies a minimal set of
characteristics, e.g., average properties
of the 96 chlorophylls,
relevant for
the description of function
and observed spectra.


The most relevant functional property of PSI
is its
high efficiency
in using excitation
energy for charge separation
in the reaction center despite thermal disorder
and ever present internal damage.
One seeks to know how {\em robust} PSI is
against perturbations such as
`pruning' or `bleaching' of individual chlorophylls
or changes in the protein
that effect chlorophyll site energies.
Robustness in biology is not a new concept.
The issue was addressed earlier in
Refs.~\cite{PRIG77,EIGE77}
and, recently in the study of  biochemical
networks involved in bacterial chemotaxis~\cite{LEIB97,LEIB99}.
These studies suggest that
biological networks
exhibit a high degree of
insensitivity to fine-tuning of physical parameters.

Another issue of interest is the {\em optimality}
of the design of the chlorophyll aggregate in PSI.
Is it possible to rearrange or reorient
the pigments in  PSI to achieve a better quantum
yield
or did evolution establish an optimal
arrangement?
This is likely easier to answer {\em in silico}
than {\em in vivo}.

While the issues of robustness and optimality
can be addressed
well by the physicists' approach,
details of the low temperature
excitation transfer dynamics and
spectral features
are likely better studied by the
chemists' approach.
The longest wavelength features of the
spectrum attributed to so-called `red chlorophylls'
are an example of this.
Red chlorophylls are responsible
for broadening the absorption profile
of PSI, but
their role is not fully understood.
For a review we refer the reader to Ref.~\cite{GOBE2001}.

In order to
answer the questions
raised above
we study ensembles of various
chlorophyll arrangements.
This permits us to investigate the
dependence of the light harvesting function of PSI in
detailed or coarse descriptions as well
as the robustness and optimality
of the system.
For this purpose,
we use the high
resolution structure
reported in~\cite{JORD2001}
as a starting point
to describe the spectral and excitation transfer
characteristics of PSI.
The structure
reveals the relative orientation
of chlorophyll molecules,
which makes
it possible to construct
an effective Hamiltonian
for the chlorophyll aggregate.
The resulting quantum yield
for charge separation
and the average
excitation lifetime is determined
based on this Hamiltonian.

We also investigate the excitation transfer
dynamics that results after the absorption
of a photon by PSI. Besides
determining the overall lifetime of an
excitation and the quantum yield
(the probability that photon absorption
leads to electron transfer from the
special pair), one wishes to know how
excitation migrating through PSI
repeatedly reaches
and escapes the
special pair, i.e., engages in
sojourns from the special pair into
the chlorophyll pool.
We suggest a kinetic
expansion method, the sojourn expansion,
to account for the outlined
excitation migration in a light-harvesting
complex.

The organization of this paper is as follows:
In the next section we introduce the effective
Hamiltonian for the chlorophyll
aggregate of PSI.
Sec.~3 utilizes this Hamiltonian
to discuss candidates for the red chlorophyll
states.
In Sec.~4 we discuss the effect of thermal disorder
on the spectral properties of PSI in terms of random
matrix theory.
In Sec.~5 we introduce
the general theoretical framework for the
study of excitation transfer dynamics,
including the sojourn expansion,
 and apply it to PSI.
The robustness of the efficiency
of the system is investigated
under the effects
of the fluctuations in chlorophyll
site energies and
the pruning
of individual chlorophylls.
In Sec.~6 we examine the optimality of the arrangement
of chlorophylls in PSI.
Sec.~7 contains a general discussion and our conclusions.
In an appendix we summarize our
method to compute inter-chlorophyll couplings.

\section{Effective Hamiltonian for chlorophyll
aggregate}

In this section we construct an
effective Hamiltonian for the aggregate of chlorophyll molecules in
PSI defined in a basis of
single chlorophyll excitations.
Isolated cyanobacterial PSI is known to
exist {\em in vivo} mostly as a trimer~\cite{JORD2001}.
In this paper we
will examine the behavior of the monomeric complex.
Of the 96 chlorophylls present in PSI, 95 are chlorophyll $a$ molecules
and one of the special pair chlorophylls is a chlorophyll  $a'$.
For computational simplicity, we will assume below all chlorophylls to be
of chlorophyll $a$ type.

The lowest excited
state of a chlorophyll molecule is the $Q_y$ state \cite{SCHE91B}. A set
of basis states for an effective Hamiltonian is then given by
\begin{equation}
|i \rangle=|\phi_1\: \phi_2\:\cdots
\phi^*_{i}\:\cdots\: \phi_{N}\rangle,
\;\;\;\;\;\; i=1,2,...N,
\end{equation}
in which the
$i$th chlorophyll is $Q_y$-excited, all the other chlorophylls being
in their ground states.
The total number of chlorophylls is $N=96$.
By using this basis, an
effective Hamiltonian can be written as
\be
H_0 = \left( \ba{ccccc}
\epsilon_1 & W_{12} & W_{13}  & \cdots & W_{1N} \\
W_{21} & \epsilon_2 & W_{23}  & \cdots & W_{2 \, N} \\
\cdots & \cdots & \cdots & \cdots & \cdots \\
W_{N1} & W_{N \, 2} & W_{N \, 3}
& \cdots & \epsilon_N \ea
\right).
\label{H0def}
\ee
Here, $\epsilon_i$ denotes the chlorophyll site energy
for the $Q_y$ state of the $i$th chlorophyll, which needs
to be determined later, and $W_{ij}$ denotes the
coupling between the $i$th and the $j$th
chlorophylls.

The electronic coupling, $W_{ij}$, between
excited states of two chlorophyll molecules has two contributions
\begin{equation}
W_{ij}
= W^c_{ij}
+
W^{ex}_{ij},
\label{Wclex}
\end{equation}
where $W^c_{ij}$ corresponds to a direct Coulomb term
\cite{FORS48} and $W^{ex}_{ij}$ corresponds to an electron
exchange term \cite{DEXT53}. In PSI, almost all of the chlorophylls are
at a Mg-Mg distance of greater than 7~\AA, a range in which
Coulomb interaction dominates~\cite{DAMJ99};
therefore we keep only the
contribution of the direct Coulomb term, $W^c_{ij}$,
in the remainder of the paper.

The electronic couplings
due to the direct Coulomb interaction
are calculated
through two approaches.
In the first (approximate) approach,
the couplings are determined
assuming a dipole-dipole
interaction between transition dipole
moments of the chlorophyll $Q_y$ states.
In this approximation, the couplings are
given by (see, for example, Ref.~\cite{DAMJ99})
\be
W_{ij} = C \left(
\frac{{\bf d}_i \cdot {\bf d}_j}{r_{ij}^3} -
\frac{3 ({\bf r}_{ij} \cdot {\bf d}_i)
 ({\bf r}_{ij} \cdot {\bf d}_j) }{r_{ij}^5}
\right)
\label{defWij}
\ee
where ${\bf d}_i$ are unit vectors along transition
dipole moments from the ground state to the $Q_y$ state of the
$i$th chlorophyll and ${\bf r}_{ij}$ is the
vector connecting
the Mg atoms of
chlorophylls $i$
and $j$;
the transition dipole moments
for the $Q_y$ state are taken to be along the vector
connecting the N$_B$ and N$_D$ atoms.
The latter assumption is seen to be valid
by comparing the transition dipole moment
unit vectors thus computed to the ones
obtained in the course of the full Coulomb
semi-empirical method introduced below.
The atomic positions are
taken from crystallographic data \cite{JORD2001}. The parameter
$C = 116 000 \, {\rm \AA}^3 \, {\rm cm}^{-1}$
in Eq.~(\ref{defWij})
is determined by
enforcing asymptotic agreement at large distances
with the full Coulomb couplings
introduced below~\cite{DAMJ99}.
Full Coulomb couplings in turn are fixed
by the oscillator
strength of the $Q_y$ transition, which is taken to be
$23$~Debye$^2$ \cite{DURR95}. Since the
transition dipole moment vectors are computed
simply from geometry,
application of Eq.~(\ref{defWij}) has
a low computational cost.
Naturally, the dipolar approximation is
the more accurate, the larger the distance between
the chlorophylls.

The second approach, following Ref.~\cite{DAMJ99},
computes
electronic couplings using a semi-empirical
Pariser-Parr-Pople (PPP) Hamiltonian
as utilized in Refs.~\cite{DAMJ99,DAMJ2000A}.
In this approach, the contribution of all
orders of multipoles are included, i.e., not only
the dipole-dipole contribution.
Therefore, we will refer to this description as
the full Coulomb (FC) approach.
The details of this approach are
outlined in the appendix.

In Fig.~\ref{fig:ps1_eigs} are illustrated
the first five eigenstates of the effective
Hamiltonian~(\ref{H0def}) with full Coulomb couplings
and identical site energies. Excitons are localized
typically only over a few chlorophylls,
in some cases coinciding with strongly
coupled pairs.

\section{Red chlorophyll candidates for photosystem I}

A number of chlorophyll molecules in PSI of {\em S. elongatus}
are known to absorb light at longer wavelengths
than the special pair P700. The number of these `red chlorophylls'
is suggested to be between seven
and eleven~\cite{PALS98,ZAZU2002}.
In Ref.~\cite{ZAZU2002} individual absorption peaks
for the red chlorophyll states
were observed for 4~K at wavelengths 708~nm, 715~nm, and 719~nm.
Among possible causes for this red shift in the
absorption peak of these chlorophylls are
the effect of the local protein environment on the site energies
and the effect of
excitonic splitting due to a strong coupling
between pairs of chlorophylls.

\begin{figure}
\centerline{\includegraphics[width=8.25cm]{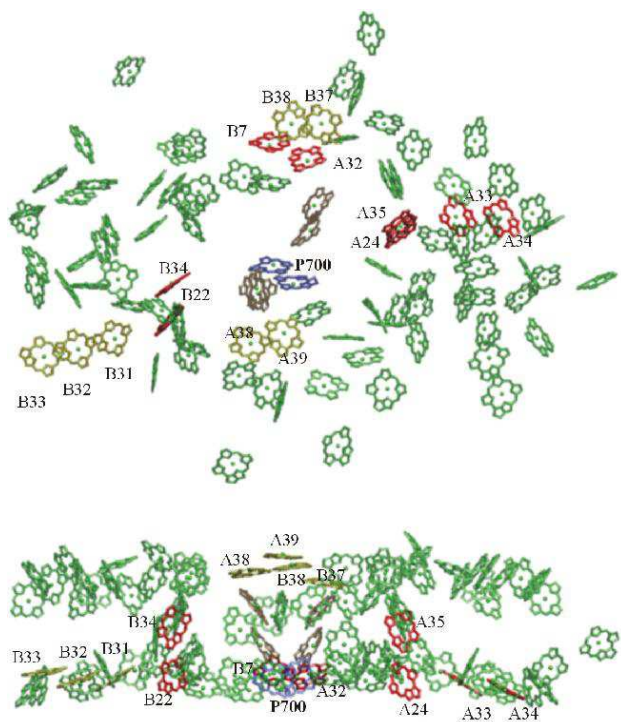}}
\caption[ps1fig]{ Red chlorophyll candidates.
Chlorophylls suggested to be engaged
in red absorption are highlighted
in red.
Red chlorophyll candidates are
chosen to be the four chlorophyll pairs,
which exhibit the strongest couplings
in the full Coulomb description
as well as a higher oscillator strength
in the lower lying excitonic state.
In yellow are shown chlorophyll pairs
with strong couplings in the dipolar approximation
that are suggested
to be {\em not} responsible for the red absorption
in the full Coulomb description
(see also text, Table \ref{tbl:chl_pairs} and
Fig.~\ref{fig:trimer}).
Figure produced with VMD~\cite{HUMP96}.}
\label{fig:ps1_redchl}
\end{figure}

In Ref.~\cite{JORD2001} candidates
for the red chlorophylls
were suggested based on the orientation of
chlorophylls and the assumption of a strong excitonic splitting
as the cause of the spectral red shift.
In Ref.~\cite{ZAZU2002} a match between these candidates
and the experimentally resolved red absorption peaks
were suggested.
In this section we extend this discussion by comparing the
dipolar couplings with the corresponding full Coulomb
couplings. In Ref.~\cite{JORD2001} a trimer
(B31-B32-B33) and three dimers (A32-B7, B37-B38, A38-A39) of
chlorophylls were given as red chlorophyll candidates
(see Fig.~\ref{fig:ps1_redchl};
nomenclature of Ref.~\cite{JORD2001}
is used for naming chlorophylls).
Indeed in the framework of the dipolar couplings these
chlorophylls form the strongest coupled pairs. However, in
the framework of the full Coulomb couplings a different
picture emerges.
This may not be surprising as the two approaches tend to
disagree over short distances, such as the
inter-chlorophyll distances within
strongly coupled pairs.

In the full Coulomb description the
couplings of the chlorophyll pairs in the trimer B31-B32-B33
are significantly (more than a factor 3) smaller than their
corresponding dipolar couplings.
Since there are chlorophyll pairs with much
stronger couplings in PSI,
the trimer loses its position
as a red chlorophyll candidate in the
full Coulomb picture.

Furthermore, a strong coupling is not enough
to cause a red shift in the absorption spectrum.
Viewed as a two state system, a chlorophyll pair
will have one higher energy excitonic state and
one lower energy excitonic state, each with a
certain oscillator strength.
A red shift
can be associated only with a case where
the lower energy excitonic state
has a higher oscillator strength than the
higher energy excitonic state.
Otherwise, there needs to
be a significant shift in the chlorophyll site energies
due to the protein environment to account for the
red shift.
It turns out that the two strongest coupled pairs,
B37-B38 and A38-A39, have most of their oscillator
strength in the higher energy state. Therefore, we
leave them out of the list of red chlorophyll candidates.

Four pairs of chlorophylls,
which have the strongest couplings
as well as the
higher oscillator
strength in the lower lying state,
are suggested as red chlorophyll
candidates (highlighted in Table \ref{tbl:chl_pairs}
and Fig.~\ref{fig:ps1_redchl}).
A match between the observed red absorption
peaks and the suggested candidates cannot
be done reliably without knowledge of the
site energies. However, matching the furthest
red shift with the strongest coupling
we tentatively suggest that A32-B7 is
responsible for the 719~nm absorption peak,
A33-A34 and A24-A35 for the 715~nm peak, and
B22-B34 for the 708~nm peak.

It must be noted
that  excitonic splitting alone cannot be
the only cause
for observed red shifts. For example,
in case of two identical chlorophylls, which have their
$Q_y$ absorption peaks at 675~nm, a coupling
of about 900~cm$^{-1}$ would be needed
to place the lower energy excitonic state
at 719~nm. This suggests that the red chlorophyll
states are caused by the contribution
of both excitonic splitting and
site energy shifts due to the local
protein environment. Furthermore, any isolated
chlorophyll may contribute to the red chlorophyll
band of the PSI spectrum solely due to its site energy shift,
a possibility which is totally neglected in
our present picture.

\begin{figure}
\centerline{\includegraphics[width=8.25cm]{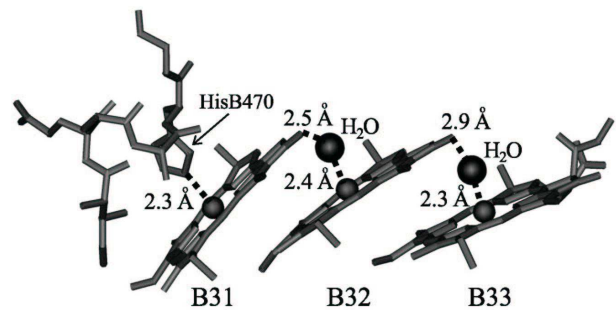}}
\caption[ps1fig]{  Chlorophyll trimer B31-B32-B33.
Shown are these three
closely spaced chlorophylls
connected
by two water molecules and coordinated
by a His residue.
The structure suggests that the trimer be
described
as a supermolecule.
Figure produced with VMD~\cite{HUMP96}.
}
\label{fig:trimer}
\end{figure}

The trimer
of chlorophylls, B31-B32-B33, deserves
a closer inspection than presented
thus far. As seen in Fig.~\ref{fig:trimer},
the atomic structure indicates that
the three chlorophylls are
connected to each other by water molecules,
suggesting that they might have to be treated
computationally as a supermolecule.
In regard to the possible failure of
our description, we note that
the His residue coordinating
chlorophylls B31-B32-B33 is
absent in {\em Synechocystis}, which
exhibits a weaker long wavelength absorption band.
Indeed, the close packing of
chlorophylls B31-B32-B33
is reminiscent
of J-aggregates~\cite{MAY2000}, which are known to have
red shifted spectra compared to
the spectra of their building
blocks.
We may suggest, hence, that the
chlorophylls B31-B32-B33
may yet be
responsible for red absorption,
despite
the weak couplings emerging
from our description.

\section{Effect of thermal disorder on spectral properties}

Spectral properties and excitation transfer dynamics
of a photosynthetic system
are two complementary sources of information.
Both are determined by the
Hamiltonian describing the system.
However, the spectral properties are
influenced more by chlorophyll site energies,
i.e., the diagonal entries of the effective Hamiltonian,
whereas
the excitation transfer dynamics more directly
reflects the role of inter-chlorophyll couplings.

A reconstruction of the (low temperature) absorption spectrum
from the Hamiltonian is not possible without
a knowledge of the site energies.
This is seen, for example, in the
much smaller seven chlorophyll system of the FMO
complex \cite{LOUW97}.
Similarly the inverse problem of using
the spectrum to reconstruct the site energies
(assuming the couplings are known) is ill posed,
since with as many parameters as the number of chlorophylls
in the system one cannot unambiguously
assign the site energies. Indeed it is a very
important test for any calculation of the
chlorophyll site energies to faithfully
reproduce the experimentally observed
absorption spectrum, especially at low temperatures.

Nevertheless spectral data can provide valuable
information on the general distribution
of chlorophyll site energies if not on individual sites.
Therefore, we study the spectral
properties of PSI in the light of the effective
Hamiltonian introduced in Sec.~2,
before we proceed with a detailed discussion
of excitation transfer dynamics.

\subsection{Random matrix theory as a description of
thermal disorder}

Both the spectral features and the excitation transfer
dynamics in a photosynthetic system need to be described
in the presence of strong thermal disorder.
Standard zero-temperature quantum theory remains
inadequate in this respect.
Below we present a method,
based on random matrix theory
\cite{PORT65,MEHT91,GUHR98},
to study static disorder
in a system described by an effective Hamiltonian,
such as the one given in Sec.~2.
For a more detailed account of the
application of random matrix theory to
the description of static disorder in
photosynthetic systems the reader
is referred to Ref.~\cite{SENE2001}.
For a discussion of dynamic disorder
in a light-harvesting complex the reader
is referred to Refs.~\cite{DAMJ2001,BARV99}.

We consider an ensemble of Hamiltonians that
describes a set of photosynthetic systems in thermal equilibrium.
In contrast to a canonical ensemble from statistical mechanics,
whose members are points in the phase space of one Hamiltonian,
we consider an ensemble of Hamiltonians each of which
represents another copy of the photosynthetic system in
question. We will be interested in spectral properties
as averaged over this ensemble.

For a description of this ensemble, we employ the
sum of the noise-free effective Hamiltonian, $H_0$,
based on the crystal structure,
and a random part, $R$, representing thermal fluctuations
\be
H = H_0 + R.
\label{Htot}
\ee
Here the matrix $R$ is drawn from
a certain probability distribution, $P(R)$,
representing the effects of thermal disorder.
All relevant spectral quantities
are defined in terms of an ensemble average
over the random part, $R$, with respect to the
weight function, $P(R)$.

The density of states and the directionally averaged
absorption spectrum for (\ref{Htot})
are defined by
\be
\rho(\omega) & = & \left<
\sum_{i=1}^{N} \delta(\omega - E_i)
\right>,
\\
\alpha(\omega)  & = &
\frac {4 \pi^2 \omega n} {3 c}
\left<
\sum_i \left| {\bf D}_i^2 \right| \delta(\omega - E_i)
\right>,
\label{alphadef}
\ee
where $<\ldots> \equiv \int d[R] P(R) \ldots \; $
and $E_i$ are the eigenvalues of $H=H_0+R$.
The transition dipole moments for eigenstates,
${\bf D}_i = \sum_m c_i(m) {\bf d}_m $,
are given in terms of the transition dipole
moment unit vectors of individual sites,
${\bf d}_m$, and the expansion coefficients, $c_i(m)$,
of eigenstates.

These definitions for the average spectral behavior
are superior to the `Gaussians on sticks' approach
sometimes utilized,
where Gaussian envelopes are put around the eigenvalues
of $H_0$ to reconstruct the spectrum.
Especially in cases where there is a significant
overlap between the envelopes of multiple eigenvalues,
the average spectral properties are seen to deviate
significantly. As an extreme example one can consider
a case of $N$ degenerate eigenvalues~\cite{MEHT91}.
In this case a sum of Gaussian envelopes
gives another Gaussian with the same width, whereas
the width of the average spectrum should be
proportional to $\sqrt{N}$.

The nature of the probability distribution, $P(R)$,
as a function of temperature is not easy to determine.
However, studies in random matrix theory have shown that
many average spectral features are largely
independent of the exact shape of $P(R)$ and
instead depend mainly on its width~\cite{GUHR98}.
This kind of independence from the
specific probability
distribution is akin to the central limit
theorem and is known as random matrix universality.
This universality allows one to make
simplifying assumptions on the nature
of $P(R)$~\cite{GUHR98}.

Although most results on random matrix universality
are based on the limit where the
matrix size becomes large, remarkable spectral similarity
persists even for mesoscopically sized ensembles.
In an earlier investigation, we have applied
random matrix universality
to the peripheral light-harvesting complex,
LH2, of purple bacteria~\cite{SENE2001}, which has 16
strongly coupled
bacteriochlorophylls. It was seen that, when the width
of the disorder term, $P(R)$, is taken
into account the average density of states and
the absorption spectrum
are rather insensitive to the
changes of $P(R)$.
Analytical formulae
for the density of states at finite temperature
as a function of the zero temperature spectrum
were derived for a simple form of $P(R)$
in Ref.~\cite{SENE2001}.

\begin{figure}
\centerline{\includegraphics[width=8.25cm]{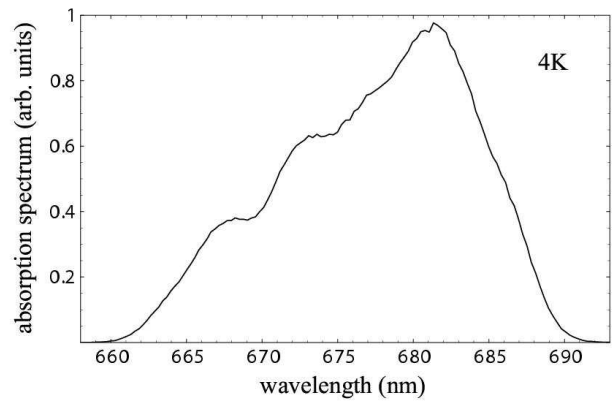}}
\caption[ps1fig]{ Absorption spectrum.
Shown is the low temperature (4~K)
spectrum for an
effective Hamiltonian with full Coulomb couplings
and homogeneous site energies (675~nm).
The spectrum is generated for an ensemble with
diagonal disorder corresponding to 70~cm$^{-1}$.
A comparison of the width of this spectrum with
that of the experimental absorption
spectrum~\cite{ZAZU2002}
allows the prediction
of the heterogeneity in the site energies.
}
\label{fig:absspec_hom}
\end{figure}

In the following we take the probability
distribution, $P(R)$,
to describe diagonal disorder
\be
P_{diag}(R)
&=& N_{diag} \;
\prod_i \exp \left(
- \frac 1 {2  \nu_{diag}^2} \;  R_{ii}^2
\right)
\prod_{i \ne j} \delta(R_{ij}).
\label{PRdiag}
\ee
For the width of the distribution (\ref{PRdiag})
we take $\nu_{diag} = 70$~cm$^{-1}$
at 4~K~\cite{PIEP99A,PIEP2001}.
A similar description has been used in Ref.~\cite{HU97}
for the light harvesting system of purple bacteria.
We use the parameter $\nu_{diag}$ to numerically
evaluate the absorption spectrum
as defined in~(\ref{alphadef}).
In Fig.~\ref{fig:absspec_hom} is shown the absorption
spectrum at 4~K based on the
effective Hamiltonian with full Coulomb couplings
and homogeneous site energies corresponding to 675~nm.
Ignoring the red chlorophyll band of the spectrum,
a width for the bulk of about 230~cm$^{-1}$
(corresponding to a FWHM of 550~cm$^{-1}$)
at 4~K is reported in Ref.~\cite{ZAZU2002}.
However,
the width
for the 4~K spectrum in Fig.~\ref{fig:absspec_hom}
measures only about 140~cm$^{-1}$.
The mismatch
can be attributed to our neglect of the
site energy heterogeneity.

Though it is not
possible to reconstruct the chlorophyll site
energies from the spectrum, one can obtain
an estimate on the width of the
distribution of site energies.
In order to accomplish
this we will ignore for the moment the red chlorophyll band
of the spectrum and instead concentrate
on the main peak as given in Ref.~\cite{ZAZU2002}.
The width of the absorption
spectrum for the actual (inhomogeneous) site
energies, $W_{inhom}$, the width of the absorption
spectrum for the homogeneous site
energies, $W_{hom}$, and the width
of the distribution of site
energies (heterogeneity), $W_{\epsilon}$, are approximately related to
each other by
\be
W_{inhom}^2 = W_{hom}^2 + W_{\epsilon}^2.
\ee
As discussed above, assuming $W_{inhom} = 230$~cm$^{-1}$
and $W_{hom} = 140$~cm$^{-1}$ results in
a site energy distribution width of about
$W_{\epsilon} = 180$~cm$^{-1}$ (FWHM of 430~cm$^{-1}$).
The validity of this assertion can
be numerically verified
by generating an ensemble of Hamiltonians
with the same full Coulomb couplings but
with random site energies for the given
value of $W_{\epsilon}$.
This value
will be utilized later in the discussion
of excitation transfer dynamics.

\section{Excitation transfer dynamics in photosystem I}

\begin{figure}
\centerline{\includegraphics[width=16cm]{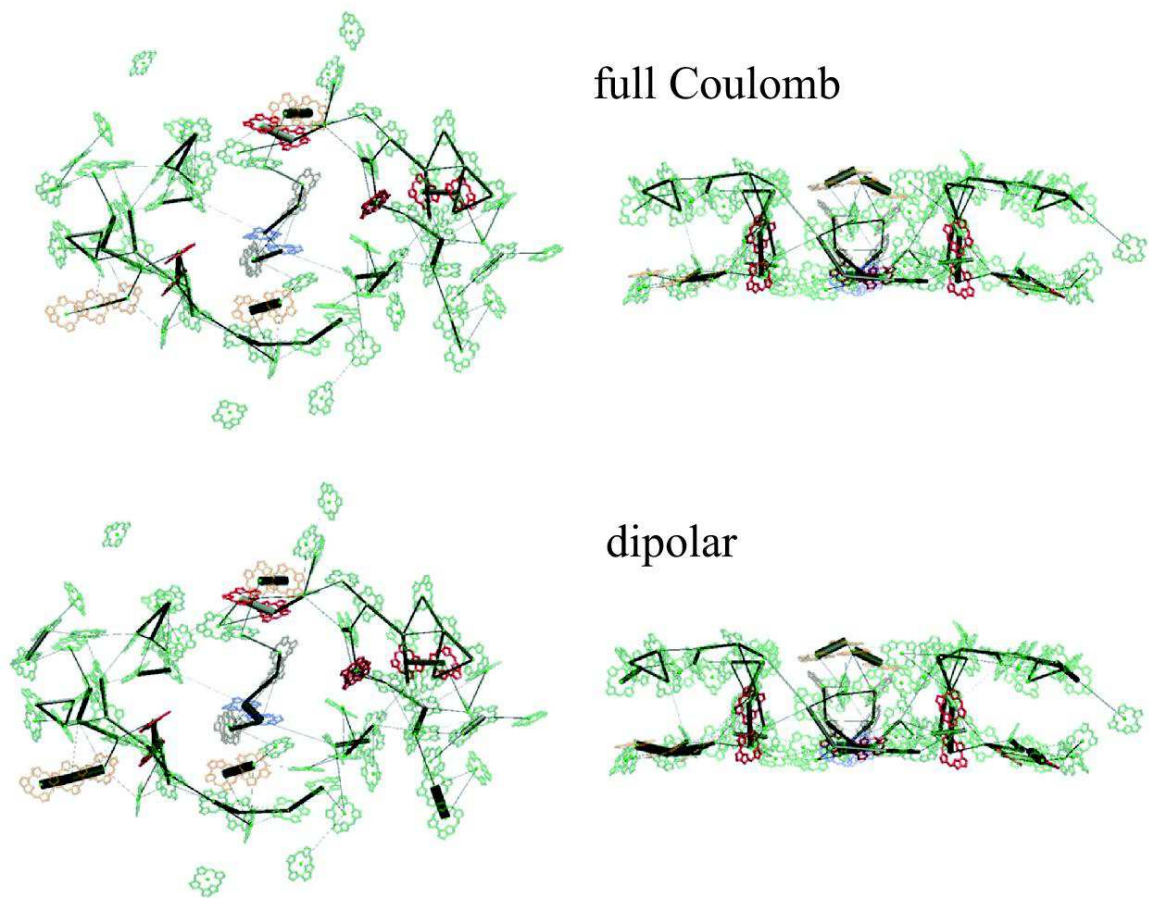}}
\caption[ps1fig]{ Network of excitation transfer
pathways in PSI.
The networks
at room temperature
for full Coulomb couplings (top) and dipolar
couplings (bottom) are shown. The thickness of  bonds
connecting chlorophylls
are proportional to the transfer rates between them.
Only the 130
largest transfer rates are presented.
The transfer rates are computed for
a system with identical site energies
at room temperature.
Figure produced with VMD~\cite{HUMP96}.
}
\label{fig:ps1_pways}
\end{figure}

In this section we construct the excitation
transfer rates between chlorophylls using F\"{o}rster
theory~\cite{FORS48}. We will follow the methodology
suggested in Ref.~\cite{RITZ2001B}.
The excitation transfer rate from chlorophyll $i$ to
chlorophyll $j$ is~\cite{FORS48,DEXT53}
\be
T_{ij}= \frac{2 \pi}{\hbar}
|H_{ij}|^2 J_{ij},
\; \; \;
J_{ij}=\int S^D_i(E) S^A_j(E) dE ,
\label{defTijJij}
\ee
where $H_{ij}$ is the coupling
between the chlorophylls and $J_{ij}$
is the spectral overlap
between the emission spectrum, $S^D_i(E)$,
of the donor chlorophyll $i$ and the
absorption spectrum, $S^A_j(E)$,
of the acceptor chlorophyll $j$.
Following Refs.~\cite{NAGA93,RITZ2001B}
we approximate these spectra
by Gaussians
\be
S^D_i(E) & = &
\frac 1 {\sqrt{2 \pi} \nu}
\exp \left( -\frac {(E_i-S-E)^2} {2 \nu^2} \right),
\el
S^A_j(E) & = &
\frac 1 {\sqrt{2 \pi} \nu}
\exp \left( -\frac {(E_j-E)^2} {2 \nu^2} \right),
\ee
where $E_i$ and $E_j$ are the absorption peaks
for the chlorophylls and $S$ is the Stokes shift.
We take the Stokes shift to be equal
to 160~cm$^{-1}$ at room temperature~\cite{SCHE91B}
and 20~cm$^{-1}$ at 4~K~\cite{PIEP2000}.
We have assumed identical widths
for the emission and absorption spectra
of 240~cm$^{-1}$ at room
temperature~\cite{SCHE91B}.
In the case
of identical site energies,
rates for
forward transfer and back transfer are
equal, $T_{ij} = T_{ji}$, but
this is not the case
when the site energies differ.
The network of excitation transfer rates
between chlorophylls of PSI for identical
site energies are shown
in Fig.~\ref{fig:ps1_pways} for
both the full Coulomb and dipolar
Hamiltonians.

\begin{figure}
\centerline{\includegraphics[width=8.25cm]{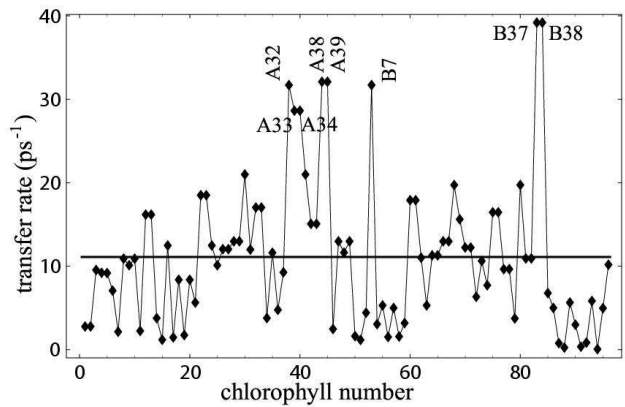}}
\caption[ps1fig]{ Largest excitation transfer rate
from PSI chlorophylls.
Shown are rates for all 96 PS1 chlorophylls
as a function of the donor chlorophyll. The horizontal line
indicates the average value of 11.1~ps$^{-1}$. The excitation
transfer rates
are computed for a Hamiltonian with full Coulomb couplings
and homogeneous site energies for
all chlorophylls, except P700 (c.f. row 1, Table~\ref{tbl:tauandq}).
}
\label{fig:maxtij}
\end{figure}

The largest transfer rate away from a given chlorophyll
suggests the average excitation lifetime
at that chlorophyll.
Fig.~\ref{fig:maxtij} shows the largest
excitation transfer rates for all 96
chlorophylls. It can be seen that these rates
correspond to an average largest transfer rate between
chlorophylls of about 11~ps$^{-1}$.

Using the transfer rates between the chlorophylls,
a master equation can be constructed
for the rate of change of probabilities,
$p_i(t)$, describing the likelihood
that the chlorophyll $i$ is electronically excited
at time t.
The corresponding rate equation is
\be
\frac {d} {d t} p_i(t)
 & = &
\sum_j K_{ij} p_j(t),
\label{defKij}
\\
K_{ij} & = &
T_{ji} - \delta_{ij} \left( \sum_k
T_{ik} + k_{diss} + \delta_{i, P700} k_{CS}
\right),
\ee
where $k_{diss}$ denotes the dissipation
(internal conversion) rate,
which is assumed to be uniform among the
chlorophylls, $k_{CS}$ denotes the charge separation
rate
at the
special pair P700 and $\delta_{i, P700}$
is equal to 1 when $i$ is one of the
two chlorophylls in the special pair and
zero
otherwise. The dissipation and charge
separation rates are not known to great
accuracy. Inspired by the purple bacterial
light-harvesting systems~\cite{RITZ2001B}
we assume a dissipation rate
of $k_{diss} = (1~{\rm ns})^{-1}$.
The charge separation rate at P700
is observed to be between
1~ps$^{-1}$ and (3~ps)$^{-1}$~\cite{BRET97,VANG94,OWEN87}.
In the following we will assume
a charge separation rate of
$k_{CS}=(1.5~{\rm ps})^{-1}$.

Given an initial distribution, $p_i(0)$,
the formal solution to Eq.~(\ref{defKij})
can be written
\be
\left| p(t) \right>
= \exp(K t) \left| p(0) \right>.
\label{poftsoln}
\ee
Using this expression, explicit formulae
for
the average excitation lifetime,
the quantum yield, and overall dissipation
rate can be given~\cite{RITZ2001B}.

Let us define a uniform (non-normalized)
state, $ |{\bf 1}> \equiv  \sum_i |i>$,
describing equally likely occupation
probabilities for all chlorophylls.
If we assume this uniform
distribution for the initial state, i.e.,
$\left| p(0) \right> = N^{-1} |{\bf 1}>$,
then the average
excitation lifetime, $\tau$, the quantum
yield, $q$, and the overall dissipation
probability, $d$, are all given in terms
of similar expressions~\cite{RITZ2001B}
\be
\tau & = &
- \frac 1 N \left< {\bf 1} \right | K^{-1} \left| {\bf 1}\right>,
\label{deftau}
\\
q & = &
- \frac 1 N k_{CS}
\left< P700 \right| K^{-1} \left| {\bf 1} \right>
= 1 - d,
\label{defqyield}
\\
d & = &
- \frac 1 N k_{diss}
\left<{\bf 1}\right | K^{-1} \left| {\bf 1} \right>
= k_{diss} \tau,
\ee
where $\left| P700 \right> \equiv \sum_i \delta_{i, P700}
\left| i \right>$.

Average excitation lifetimes
and the quantum yields for various models
are provided in Table~\ref{tbl:tauandq}.
The data in the first three rows are based on
an effective Hamiltonian that
assumes identical site energies
for the two chlorophylls of the P700 pair.
The site energies were chosen such that
the P700 excitonic state
with the highest oscillator strength
coincides with the 698~nm (14327~cm$^{-1}$) absorption peak at 4~K.
Since one of the two P700 chlorophylls
is a chlorophyll $a'$, the identical site energy assumption
for the P700
is likely to be unrealistic, however
it shall be adequate for an approximate description of
the excitation transfer dynamics.
For full Coulomb couplings the highest
oscillator strength lies at the higher
energy excitonic state (see Table~\ref{tbl:chl_pairs});
since the respective
coupling between the two P700 chlorophylls
is 47.6~cm$^{-1}$, the P700 site
energy is chosen to be 14279~cm$^{-1}$.
For dipolar couplings the highest
oscillator strength lies at the lower
energy excitonic state
(the sign of the coupling is different
for the two methods);
the coupling between the two P700 chlorophylls
is 272~cm$^{-1}$
requiring a P700 site
energy of 14599~cm$^{-1}$.
The site energies for the remaining chlorophylls
are placed at 675~nm (14815~cm$^{-1}$)
unless otherwise noted.

The efficiency of 97.3~\%
and the average excitation lifetime
of 27.4~ps for
full Coulomb couplings, given in the first
row of Table~\ref{tbl:tauandq},
are in general agreement with the near unit efficiency and
the 20-40~ps lifetime reported for PSI
at room
temperature~\cite{GOBE2001,GOBE2001A,HOLZ93,KENN2001,MELK2000,MELK2001}.
A recent estimate of the average excitation lifetime
reported for trimeric PSI in {\em  S. elongatus} is 35.8~ps~\cite{GOBE2001A}.
A comparison between the first and the last rows
in Table~\ref{tbl:chl_pairs}
shows the importance of having P700 chlorophylls
at a lower energy than the bulk of the chlorophylls
in order to decrease the trapping time
and to increase efficiency.

\begin{figure}
\centerline{\includegraphics[width=8.25cm]{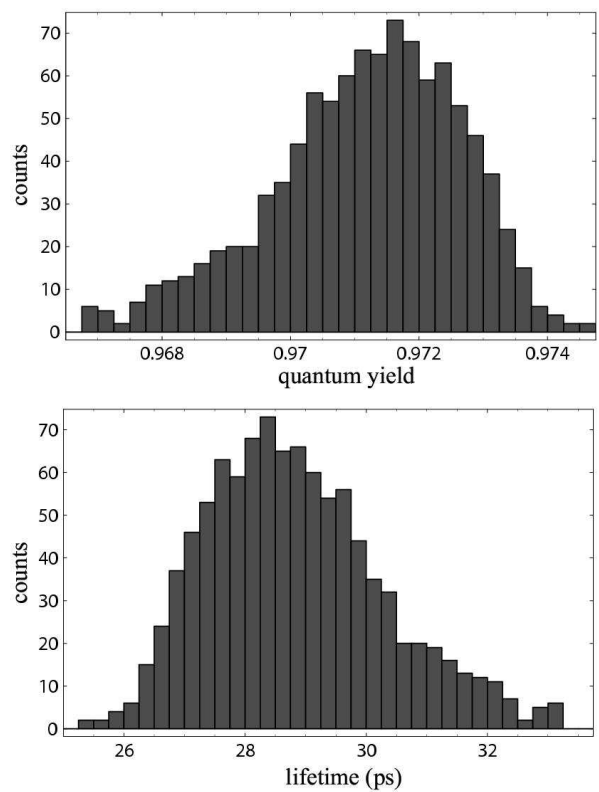}}
\caption[ps1fig]{ Effect of fluctuations
of the site energies (heterogeneity) on excitation transfer
dynamics.
Shown are the
distributions of quantum yield and
lifetime at room temperature
for an ensemble of effective Hamiltonians
with full Coulomb couplings.
The ensemble consists of systems
where P700 was tuned to the observed absorption peak
and with the rest of the chlorophyll site energies
chosen randomly with a width of 180~cm$^{-1}$
around 675~nm.
The histograms were generated
from an ensemble of 1000 systems.
}
\label{fig:randomsite_hist}
\end{figure}

In Fig.~\ref{fig:randomsite_hist}
are provided the distribution
of quantum yield and average excitation
lifetime across an ensemble of
Hamiltonians with full Coulomb
couplings, P700 tuned to the observed
absorption peak, and all other chlorophyll
site energies randomly distributed with
a width of 180~cm$^{-1}$ around 675~nm
(thus the ensemble covers different
realizations of heterogeneity, not static disorder).
Fig.~\ref{fig:randomsite_hist} reveals
that, at room temperature, fluctuations in the
site energies (heterogeneity) have no considerable
effect on efficiency. This allows
us to reconstruct room temperature
excitation transfer dynamics with
reasonable accuracy even without
detailed knowledge of the chlorophyll site energies.
However, at 4~K, the line shapes
of individual chlorophylls become so narrow that
the spectral overlap integrals in
Eq.~(\ref{defTijJij}) become unrealistically small
for misplaced site energies. Also at low
temperatures the broadening of spectral
line shapes due to electron-phonon
couplings~\cite{ZAZU2002} need to be
taken into account as this broadening is
likely to effect the overlap integrals
in the context of F\"{o}rster theory. We find it impossible
to construct a reliable picture
of the low temperature excitation transfer
dynamics with the current data.

It is of interest to know
in how far low
energy chlorophylls control the excitation transfer
dynamics and trapping in PSI.
In this regard, one may construct models
in which
the red chlorophyll candidates given in
Sec.~3 are used to refine the assignment
of chlorophyll site energies.
However, it must be kept in mind that
an accurate assignment of an individual
chlorophyll pair to an observed red chlorophyll state
cannot be done reliably without a precise knowledge
of site energies.
It is seen
that the room temperature excitation transfer dynamics is
not effected greatly by the inclusion of this extra
information in a model where all other site energies
are unknown or taken to be random.
Fig.~\ref{fig:randomsite_hist} supports this claim.
As a result we find it rather unfruitful to proliferate
the multitude of models already given in Table 2.
At room temperature the role of low energy chlorophylls
appear to be simply to extend the absorption profile to longer
wavelengths rather than to have a profound effect
on the excitation transfer dynamics.
However, at low temperatures
red chlorophylls
may function as effective traps
and could serve as probes to investigate
the excitation transfer dynamics
of PSI in detail.

\subsection{Effect of removal of individual
chlorophylls on excitation transfer dynamics}

One way to probe the function of individual
chlorophylls in PSI as well as the degree
of robustness in its design is to examine
the effect of the removal of chlorophylls
on the efficiency of the system.
To this end we remove various
chlorophylls one at a time and examine
the quantum yield of the remaining
system.

\begin{figure}
\centerline{\includegraphics[width=8.25cm]{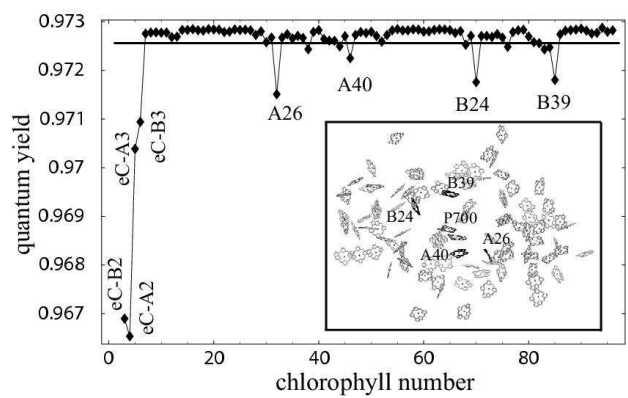}}
\caption[ps1fig]{ Quantum yield for pruned chlorophylls.
Shown is the quantum yield resulting
after pruning of individual chlorophylls.
The chlorophylls whose deletion has the highest impact
on quantum yield are labeled.
Chlorophylls (eC-xx), 
numbered 3 to 6, are part of the
reaction center. 
Chlorophylls A26, A40, B24, and B39
are connecting chlorophylls
situated closest to the reaction center (see inset).
The horizontal
line denotes the efficiency of the unperturbed
system (97.25~$\%$), which is described by the effective Hamiltonian
with full Coulomb couplings,
identical site energies, and
P700 tuned to the observed absorption spectrum
 (c.f. row 1, Table~\ref{tbl:tauandq}). }
\label{fig:qremoval}
\end{figure}

Fig.~\ref{fig:qremoval} shows the
quantum yield as a function of the removed
chlorophyll.
The two P700 chlorophylls
are not removed in this study.
Although the effect of the removal
of the other four reaction center
chlorophylls on the quantum yield are shown, it is hard
to test their effect directly,
since they are also a part of the electron transfer
chain. Therefore, we will concentrate
on the remaining 90 chlorophylls.

It is interesting to note that, except for the
four reaction center chlorophylls which are
immediate neighbors to the P700, removal of
a chlorophyll hardly has any effect on the quantum yield.
In fact, for most of the chlorophylls
a deletion results in an increase
in the quantum yield. This is a consequence
of the relative yield being defined
with respect to the remaining chlorophylls,
not with respect to the unperturbed system.

Thus it is seen that, outside the reaction
center, no chlorophylls play the role
of a gatekeeper for excitation transfer,
whose removal would have an adverse effect
on the yield.
Instead the individual chlorophylls contribute
to the overall cross-section of the antenna
complex.
Outside the electron transfer chain,
the four chlorophylls with
the highest impact on efficiency
(A26, B24, B39 and A40) are the nearest
neighbors of the six reaction center chlorophylls.
These chlorophylls, especially A40 and B39,
appear to link the antenna system to the reaction
center and are therefore referred to as
connecting chlorophylls.
As an extreme example one may consider
the removal of all four of these chlorophylls simultaneously.
Even in this case the quantum yield
reduces to only 96.85~$\%$ from
the value of 97.25~$\%$ for the
unperturbed complex with the same
Hamiltonian; the average excitation lifetime
increases from 27.4~ps to 31.5~ps
upon removal.
These results agree with some of the
simulations reported in Ref.~\cite{VALK95}.

\subsection{An expansion of the average lifetime
in terms of return times to the reaction center}

The average excitation lifetime, $\tau$, can be expanded
in terms of processes describing the initial
delivery of excitation
to the special pair and possible
subsequent returns following detrapping.
In order to establish such an expansion,
we rewrite the excitation transfer rates
(\ref{defKij}) in a form where
the P700 is treated as a single
excitonically coupled unit as opposed
to two separate chlorophylls.

A two state Hamiltonian for P700
is given by
\be
H_{P700} =
\left( \ba{cc}
\epsilon_1  & U_{P700} \\
U_{P700}  & \epsilon_2
\ea
\right),
\label{defHP700}
\ee
where $\epsilon_1$ and $\epsilon_2$
are the site energies for the two P700
chlorophylls and $U_{P700}$ denotes the coupling between
them. Not knowing the values
of the site energies it will be
assumed below that $\epsilon_1 = \epsilon_2$.
However, the expansion below can be readily
applied to nonidentical site energies.

Let us denote the eigenvalues of the two state
system (\ref{defHP700}) by $E_1$ and $E_2$
and the corresponding eigenfunctions by
\be
\Psi^{P700}_1 = c_{1,1} \left| 1 \right> +
 c_{1,2} \left| 2 \right>,
\el
\Psi^{P700}_2 = c_{2,1} \left| 1 \right> +
 c_{2,2} \left| 2 \right>,
\label{p700eig}
\ee
respectively. Then the coupling of any of the
other chlorophylls, $j=3, \cdots , N$, to any eigenstate, $m=1,2$, of
P700 is
\be
\tilde H_{mj} = \tilde H_{jm} =
\sum_{\alpha=1}^2 c_{m,\alpha} H_{j \alpha},
\ee
where $H_{j \alpha}$ are the elements of
the original effective Hamiltonian~(\ref{H0def}).

We will assume that thermal equilibration between
eigenstates of P700 have been achieved before
excitation transfer out of P700 occurs.
This is likely to be a reasonable assumption
at room temperature for full
Coulomb couplings, where $k_B T=209$~cm$^{-1}$
is larger than the interchlorophyll
couplings. The expansion outlined below
can also be formulated for a reaction
center where multiple chlorophylls
without a strong excitonic character
are responsible for charge separation,
making the thermal equilibration assumption invalid.
However, in that case the final formulation
of the expansion
cannot be written as succinctly
as Eq.~(\ref{deftausojsimple}) below.

The excitation transfer from P700 to any other
chlorophyll is given by a Boltzmann weighted sum
\be
\tilde T_{P700,j} & = & \sum_{m=1}^2 \tilde T_{mj},
\el
\tilde T_{mj} & = & \frac {2 \pi} \hbar
\frac { e^{-E_m / k_B T}} {\sum_{n=1}^2 e^{-E_n / k_B T}}
\left| \tilde H_{mj} \right|^2
\int S^D_m(E) S^A_j(E) dE,
\ee
where we assume an identical line shape for the
eigenstates of P700,
chosen equal to the lineshape of all
other chlorophylls in PSI.
The excitation transfer rate from another chlorophyll to
P700
is
\be
\tilde T_{j,P700} = \sum_{m=1}^2 T_{j,m},
\ee
where $T_{j,m}$ are given by Eq.~(\ref{defTijJij}),
except that $m=1,2$ now denotes
an eigenstate of P700 as given in Eq.~(\ref{p700eig}).

A reduced $(N-1) \times (N-1)$ transfer
rate matrix, $\tilde T$, can be constructed
using these rates.
The corresponding matrix, $\tilde K$,
that enters in Eq.~(\ref{defKij})
governing the kinetics of PSI,
is
\be
\tilde K_{ij} & = &
\tilde T_{ji} - \delta_{ij} \left( \sum_k
\tilde T_{ik} + k_{diss} + \delta_{i, P700} \, k_{CS}
\right).
\label{defredKij}
\ee
Average excitation lifetime, $\tau$,
and  quantum yield of the system are
defined as in Eqs.~(\ref{deftau})
and (\ref{defqyield}), e.g.,
\be
\tau & = &
- \frac 1 {N-1} \left< {\bf 1} \right |
\tilde K^{-1} \left| {\bf 1}\right>,
\label{defredtau}
\ee
where $\left| {\bf 1}\right> =
\left| P700 \right> + \sum_{j=3}^{N} \left| j \right>$
denotes the (nonnormalized) uniform initial state
(in the basis of this reduced Hamiltonian
$\left| P700 \right>$ denotes the vector with
a one in the first column and zeroes everywhere
else).

\vskip 0.5cm

\begin{figure}
\centerline{\includegraphics[width=8.25cm]{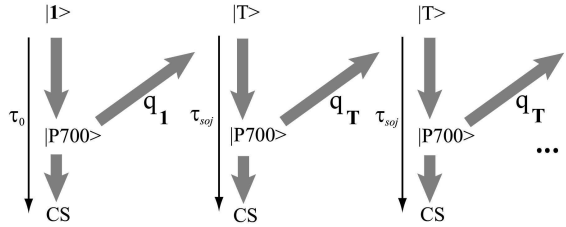}}
\caption[ps1fig]{ Processes contributing to the
overall excitation lifetime.
The excitation in a uniform initial
state
$\left| {\bf 1} \right>$  (see text)  
after reaching
the special pair $\left| P700 \right>$
either causes
a charge separation (CS)
or, with detrapping probability $q_{\bf 1}$,
returns to the surrounding chlorophylls
through the transient state $\left| T \right>$ (see text).
From the transient state excitation migrates
through PSI to return to $\left| P700 \right>$
after which charge separation occurs or, 
again with detrapping probability,
$q_{T}$, the transient state  $\left| T \right>$ 
is reached once more.
The process of migration, 
return to $\left| P700 \right>$,
possibly charge separation or renewed
detrapping is repeated.
The first usage
time, $\tau_0$, is the sum of the delivery-to-trap
time starting from a uniform state, $\left| {\bf 1} \right>$
and of the charge separation time.
$\tau_0$ represents the average time it takes
for charge separation to take place
if no detrapping events occur.
The sojourn time, $\tau_{soj}$, is the sum of the
delivery-to-trap
time starting from the transient state, $\left| T \right>$
and of the charge separation time.
After subtracting the charge separation time,
$\tau_{soj}$ may be viewed as the return time
back to P700 after a detrapping event.
Dissipation processes are not shown.
}
\label{fig:sojourn}
\end{figure}

We introduce now the expansion of the
average excitation lifetime.
The terms that arise in this expansion
are introduced and explained in
Fig.~\ref{fig:sojourn}.
For the purpose of this expansion
we separate
from  $\tilde K$
the operator, $\Delta$, that describes detrapping
\be
\tilde K & \equiv & \kappa + \Delta,
\el
\Delta & \equiv &
\left( \ba{cccc}
0 & 0  & \cdots & 0 \\
\tilde T_{P700,3} & 0 & \cdots & 0 \\
\tilde T_{P700,4} & 0 & \cdots & 0 \\
\cdots & \cdots & \cdots & \cdots \\
\tilde T_{P700,N} & 0 & \cdots & 0
 \ea
\right).
\label{defkappadelta}
\ee
The inverse, $\tilde K^{-1}$,
that enters in Eq.~(\ref{defredtau}), can  be expressed
through Taylor expansion in $\Delta$
\be
\tilde K^{-1} & = & (\kappa + \Delta)^{-1}
= \left( {\bf 1} + \kappa^{-1} \Delta \right)^{-1} \kappa^{-1}
\el
& = &
\kappa^{-1} - \kappa^{-1} \Delta \kappa^{-1} +
\kappa^{-1} \Delta \kappa^{-1} \Delta \kappa^{-1}
- \cdots
\ee
where ${\bf 1}$ is the $(N-1) \times (N-1)$ identity matrix.
The validity of this expansion can be
proven by noting that the absolute value of the eigenvalues of
$\kappa^{-1}  \Delta$ are less than one.
Substituting this expansion into Eq.~(\ref{defredtau})
results in a decomposition of the lifetime into a series
\be
\tau & \equiv & \tau_0 + \tau_1 + \tau_2 + \cdots,
\ee
where
\be
\tau_0 & = & - \frac 1 {N-1}
\left<{\bf 1}\right| \kappa^{-1} \left| {\bf 1} \right>,
\el
\tau_1 & = & + \frac 1 {N-1}
\left<{\bf 1}\right|
\kappa^{-1} \Delta \kappa^{-1} \left| {\bf 1} \right>,
\el
\tau_2 & = & - \frac 1 {N-1}
\left<{\bf 1}\right|
\kappa^{-1} \Delta \kappa^{-1} \Delta \kappa^{-1}
\left| {\bf 1} \right>,
\el
 & \cdots &
\label{tauexpand}
\ee

One can simplify Eq.~(\ref{tauexpand}) by
noting that
$\Delta$ can be expressed as
$\Delta = W_D \left| T \right> \left< P700 \right| $
where $W_D = \sum_{j=3}^N T_{P700,j}$ is the total detrapping
rate from the special pair, P700, and where
$\left| T \right> = W_D^{-1}
\sum_{j=3}^N T_{P700,j} \left| j \right> $
is a transient state representing
the occupation probabilities right
after a detrapping event has occurred.
Substituting $\Delta$, thus factored, into Eq.~(\ref{tauexpand})
the expansion
terms for the average excitation lifetime
can be written
\be
\tau_1 & = & q_{\bf 1} \; \tau_{soj} ,
\el
\tau_2 & = & q_{\bf 1} \; q_{T} \; \tau_{soj} ,
\el
\tau_3 & = & q_{\bf 1} \; q_{T}^2 \; \tau_{soj} ,
\el
& \cdots &
\label{defsojtimes}
\ee
where the defining parameters are
\be
\tau_{soj} & = &
- \left<{\bf 1}\right| \kappa^{-1} \left| T \right>,
\el
q_{\bf 1} & = &
- \frac {W_D} {N-1}
\left< P700 \right| \kappa^{-1} \left| {\bf 1} \right>,
\el
q_{T} & = &
- W_D \left< P700 \right| \kappa^{-1} \left| T \right>.
\ee
The terms appearing in Eq.~(\ref{defsojtimes})
can be summed to give
\be
\tau = \tau_0 + \frac {q_{\bf 1}} {1 - q_T} \tau_{soj}.
\label{deftausojsimple}
\ee

The expansion decomposes the average
excitation
lifetime in terms of multiple
escapes and returns (sojourns) to the
reaction center,
and hence, we call it the sojourn expansion.

Terms arising in the sojourn expansion
for an effective Hamiltonian with
full Coulomb couplings
(see row 1 of Table~\ref{tbl:tauandq})
is provided in Table~\ref{tbl:sojourn}.
It can be seen that the time
for a first delivery of excitation
to P700,
approximately given by $\tau_0$,
constitutes less than half of the average excitation lifetime
at room temperature.

\section{How optimal is the design of photosystem I?}

At first glance the arrangement of chlorophylls in PSI
appear to be random except for a pseudo-$C_2$ symmetry
(see Fig.~\ref{fig:ps1_chl_prot}).
Especially when compared to the regular
circular aggregates found in purple
bacterial photosynthetic systems,
this brings up the question of whether
the particular arrangement of chlorophylls in PSI
is optimal in some way.

Even though the fitness landscape over which
a photosynthetic apparatus should be judged
is many dimensional, a crude measure of optimality
is given by the quantum yield
of the system. Ignoring all other concerns for
optimality from an evolutionary perspective
such as assembly, aggregation
and interaction with other biomolecules in the cell,
we examine
whether the chlorophyll aggregate
in PSI can be rearranged to result in a better quantum yield.

In order to keep the number of parameters to be varied
at a manageable level, the chlorophyll
positions (as defined by their Mg atoms)
as well as the chlorophyll site energies
are kept fixed. Instead we
consider an ensemble of alternate PSI's formed
by random independent reorientations of
constituent chlorophylls.

It is costly to repeat the semi-empirical
computation of full Coulomb couplings
between chlorophylls for a large ensemble of
chlorophyll arrangements.
Therefore, we
revert to dipolar couplings for this study.
The ensemble in question is formed by
taking the vector for the transition dipole
moment at each chlorophyll and
multiplying it by a randomly
generated $SO(3)$ rotation matrix, thus
reorienting it. A new effective Hamiltonian
is then computed using the couplings
between these new
transition dipole moment vectors
as explained in Sec.~2, while
keeping the same site energies as before.
The effective Hamiltonians thus generated are
used to compute the quantum yield
and the average excitation lifetime at room temperature.
For this study identical
site energies (675 nm) for all chlorophylls
are chosen
except for P700, the site energies of which are chosen
in such a way as to coincide with the observed
absorption peak, as explained in Sec.~5.
Therefore the results of this study should be
compared with the original arrangement
of chlorophylls given in row~2 of
Table~\ref{tbl:tauandq}.

\begin{figure}
\centerline{\includegraphics[width=8.25cm]{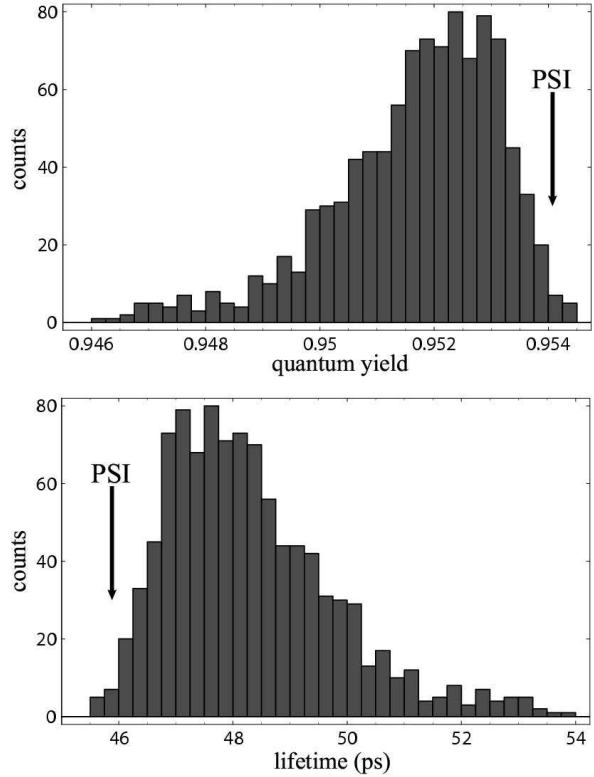}}
\caption[ps1fig]{ Optimality of PSI.
The distributions of quantum yield and
average excitation lifetime for an ensemble of
randomly generated chlorophyll aggregates
at room temperature are shown.
Chlorophyll couplings are computed in the
dipolar approximation to reduce computational cost.
P700 site energies are tuned to the observed
absorption peak and the remaining site energies are
taken to be identical (675~nm).
The histograms are generated
from an ensemble of 1000 systems.
The results for the original arrangement
(c.f. row 2, Table~\ref{tbl:tauandq})
of chlorophylls are indicated by an arrow.
}
\label{fig:randomDP_hist}
\end{figure}

Fig.~\ref{fig:randomDP_hist} contains
a histogram of quantum yields
and average excitation lifetimes for this
ensemble at room temperature. The values for the
original arrangement of chlorophylls
are indicated by an arrow. It is seen that
the original chlorophyll arrangement is
nearly at the top of the distribution, though
a few arrangements can be found to have
a better quantum yield.
It does not necessarily imply, however, that
the efficiency of PSI can be easily
improved upon. Not only
is the study of quantum yields
outlined above approximate in nature,
but also there is no way to accurately
model the constraints which had to be
satisfied over the course of the evolution
of this system. Nevertheless, it is
impressive to find the original
chlorophyll arrangement at near top efficiency.

The relatively small variation in the
quantum yield across
the ensemble seen in Fig.~\ref{fig:randomDP_hist}
brings up the question as to whether the apparent
optimality is an artifact that will disappear
in a more accurate computation.
This small variation
is a result of the
nearly three orders of magnitude difference
between the dissipation and charge
transfer rates.
As an ultimate test, full Coulomb
couplings need to be computed for each system in
an ensemble of randomly oriented chlorophyll aggregates
and the effect of individual site energies also
need to be taken into account. This is beyond
the scope of this study due to prohibitively
high computational costs.
Nevertheless, as a consistency check,
the ensemble described in Fig.~\ref{fig:randomDP_hist}
may be replaced by one with an alternate (random)
set of site energies.
It is seen that at room temperature the result of near
optimal efficiency changes
little under an assumption of different site
energies (not shown). This is because
the transfer rates at
room temperature are influenced more strongly by
couplings than by site energies, unless site
energy variations are large enough to be comparable
to the spectral width of individual chlorophylls
(see Eq.~\ref{defTijJij}).
Also, the similarity
between the networks of excitation transfer pathways
for dipolar and full Coulomb couplings seen in
Fig.~\ref{fig:ps1_pways} seems to suggest that
the general properties of excitation transfer dynamics
and therefore also the optimality
will not be greatly effected by going from the
dipolar to the full Coulomb couplings.

\section{Conclusion}

The availability of a high resolution
structure of the cyanobacterial photosystem~I
provides new
opportunities to investigate the mechanism of
light harvesting in this large pigment-protein complex.
In this study,
the excitation transfer dynamics
at room temperature
has been examined
starting from an effective
Hamiltonian based on interactions between
chlorophylls.

As a first approximation
the effect of chlorophyll-protein
interactions on the site energies
has been ignored.
However, we have used the available spectral data
to construct a more refined Hamiltonian
with limited information on site energies.
Also, a prediction on the overall heterogeneity of
site energies is made.

The calculated quantum yield and  average
excitation lifetime at room temperature
compare favorably with experimental results.
The sojourn expansion method
resolves details of excitation
migration in PSI, namely
repeated approaches to and
escapes from the reaction center.
Studying an ensemble of Hamiltonians
corresponding to copies of PSI with varying
site energies,
it is seen that the yield and the lifetime are
influenced largely by inter-chlorophyll
couplings and not so much by fluctuations of the
site energies at room temperature.
The sojourn expansion reveals that the
time for a first delivery of excitation
to P700
contributes
less than half of the average excitation
lifetime.
At room temperature,
about two thirds of the
lifetime results from detrapping events.

However, the low temperature behavior
of PSI is more difficult to describe
accurately than the room temperature
behaviour. Not only is the low
temperature absorption
spectrum impossible to reproduce without
a knowledge of chlorophyll site
energies, but also overlap integrals
contributing to the excitation transfer
rates become much more sensitive to the
difference between the donor and the acceptor
site energies due to the smaller linewidths
at low temperatures.
Furthermore, the
broadening of lineshapes due
to electron-phonon couplings also need to be
carefully taken into account at low
temperatures as this effect may have a significant
impact on the overlap integrals in the framework of
F\"{o}rster theory. Ignoring these contributions
results in an unrealistic picture of
excitation transfer dynamics at low temperature.

The robustness and optimality of the chlorophyll arrangement
in PSI has been examined by means of various
computational experiments. On the one hand, it is
seen that the efficiency of the chlorophyll
network of PSI is robust against
perturbations such as the pruning of individual
chlorophylls or fluctuations of
chlorophyll site energies.
On the other hand,
a study of an ensemble of alternate PSI's with
randomly reoriented chlorophyll aggregates
shows that the original arrangement actually has
a near optimal quantum yield.

\section*{Note added in revision}

After submission of this paper we have been
informed of a recent calculation of chlorophyll site energies
for cyanobacterial PSI~\cite{DAMJ2002}.
When these site energy values are substituted
in the computations outlined above we obtain
at room temperature
an average excitation lifetime of 33.9~ps
and a quantum yield of 96.6~$\%$, which compare
favorably with the results given in this paper.
The overall
width of the distribution of site energies,
including the red chlorophylls and P700,
is around 260~cm$^{-1}$, which is larger
than the estimate made in this paper of
180~cm$^{-1}$ based on the bulk of the low temperature
spectrum.
These new results support
our suggestion that
the room temperature excitation transfer dynamics of PSI
can be described well without
detailed knowledge of site energies.
Repeating the optimality study discussed in section 6
also for 
the set of site energies in Ref.~\cite{DAMJ2002}
yields results similar to those shown 
in Fig.~\ref{fig:randomDP_hist}.

\section*{Acknowledgements}

The authors would like to thank
A. Damjanovi\'{c}, J. Gullingsrud, S. Hayashi,
R. Knox, G. Small, E. Tajkhorshid, and R. van Grondelle
for useful discussions.
This work has been supported by NIH grants
PHS 5 P41 RR05969 and PHS 1 R01 GM60946
and the Fetzer Institute.

\appendix

\section{Full Coulomb computation of
inter-chlorophyll couplings}

Here we outline the method employed
to compute the electronic couplings, $W_{\alpha\beta}$,
between two chlorophylls semi-empirically.
This section follows Ref.~\cite{DAMJ99}
to which the reader is referred  for
a more detailed treatment.

The electronic coupling between a donor
and an acceptor molecule can be written 
\begin{equation}
W_{\alpha\beta}=\sum_{i,j\in I_D}\sum_{R,S\in I_A}
 C^c_{ij,RS}\times\langle\psi^*_{\alpha}|^{00}O^i_j|\psi_{\alpha}\rangle
\times\langle\psi_{\beta}|^{00}O^R_S|\psi^*_{\beta}\rangle,
\end{equation}
where $I_D$ and $I_A$ denote the set of atomic orbital
indices of the donor and acceptor chlorophylls,
and  $C^c_{ij,RS}$ describe the Coulomb integrals
involving atomic orbitals labeled by $i,j, R$ and $S$.
The spin tensors $^{00}O^i_j$ and $^{00}O^R_S$ prompt
the intramolecular transitions
$|\psi_{\alpha}\rangle \rightarrow |\psi^*_{\alpha}\rangle$,
$|\psi^*_{\beta}\rangle \rightarrow |\psi_{\beta}\rangle$.
The Coulomb interaction has zero rank, as denoted by the
$00$ superscript.
Therefore it only proceeds through singlet-singlet transitions.
The Coulomb integral
$C^c_{ij,RS}$ can be approximated \cite{NAGA93} as
$S_{ij}\cdot \frac{e^2}{R_{ij,RS}}\cdot S_{RS}$,
where $S_{ij}$ and $S_{RS}$ are the atomic-orbital overlap
integrals, and $R_{ij,RS}$ is the distance between
the midpoint of atoms $i$ and $j$ and the midpoint of the
atoms $R$ and $S$. As suggested in \cite{WEBE96},
$S_{ij}$ is taken to be $1$ when $i=j$,  0.27 when atoms
$i$ and $j$ are joined by a chemical bond, and zero otherwise.

To evaluate the transition density matrix elements, one requires
the description of the chlorophyll electronic states
involved in the excitation
transfer process. We choose a semi-empirical description for the
chlorophyll electronic states as provided by the
Pariser-Parr-Pople (PPP)
Hamiltonian \cite{PARI53,POPL53}, which has been used earlier in
\cite{DAMJ99,DAMJ2000A}.  The PPP Hamiltonian
\begin{eqnarray}
H_{PPP} &=&
\sum_{i<j}Z_iZ_jR_{ij}+\sum_{i,\sigma}
\left( -I_i-\sum_{j\ne i}Z_jR_{ij}
\right)
n_{i\sigma} \nonumber \\
&+&\sum_{i\ne j,\sigma}
t_{ij}c^+_{i\sigma}c_{j\sigma}+\frac{1}{2}
\sum_{i,j,\sigma,\tilde{\sigma}}
R_{ij}n_{i\sigma}n_{i\tilde{\sigma}}
\label{eq_PPP1}
\end{eqnarray}
involves orbitals of $\pi$-type only.
The creation and annihilation operators,
$c^+_{i\sigma}$ and $c_{j\sigma}$,
act on mutually orthogonal atomic $\pi$-orbitals.
The corresponding
number operator is given by $n_{i\sigma}=c^+_{i\sigma}c_{i\sigma}$.
$R_{ij}$ is the effective electron-electron repulsion integral
between an electron in atomic obital
at site $i$ and one in orbital at site $j$ and
$t_{ij}$ denotes the resonance integral between
atoms $i$ and $j$. $I_i$ is
the effective ionization potential of an orbital at site $i$.
$Z_i$ is the net charge of the core at atom $i$ which
was chosen to be 1.
The resonance integral, $t_{\alpha\beta}$,
is evaluated according to the empirical formula~\cite{SCHU76}
\begin{equation}
t_{\alpha\beta}=
\gamma_0+3.21(r_{ij}-1.397{\rm \AA}),
\label{eq_PPP2}
\end{equation}
where $\gamma_0$ is a constant and
$r_{ij}$ is the distance between the nuclear sites $i$ and $j$.
The effective electron-electron repulsion
integral, $R_{ij}$, is calculated through the
Ohno formula~\cite{TAVA86A,TAVA87},
\begin{equation}
R_{ij}=14.397\; eV
\left [\left( \frac{2\times14.397\; eV}
{R_{ii}+R_{jj}}\right)^2+
\frac{r^2_{ij}}
{{\rm \AA}^2}\right]^{-\frac{1}{2}}. \label{eq_PPP3}
\end{equation}
Assuming the same values of the semiempirical parameters as in
Ref.~\cite{DAMJ99} (listed in Table~\ref{tbl:fcparams}), we perform a
SCF-CI calculation, including all single excited $\pi$-orbitals,
for the singlet states of a chlorophyll.

\begin{figure} 
\centerline{\includegraphics[width=8.25cm]{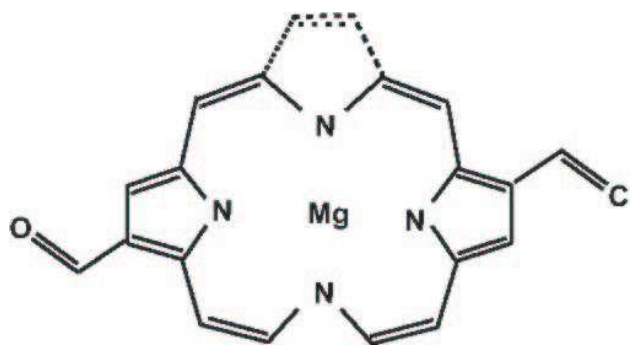}}
\vskip 4mm \caption{ Schematic representation of the conjugated
double bonds of the chlorophyll analog (in solid line) used in the
calculations of the couplings.
Part of ring II (dotted line)
has been removed
from the structure (see text).}
\label{fig:chlanalog}
\end{figure}

The chlorophyll electronic structure
calculations are based on geometries
of a simplified chlorophyll analog,
displayed in Fig.~\ref{fig:chlanalog} (solid line).
In this analog, the
double bond of ring II (dotted line
in Fig.~\ref{fig:chlanalog})
is taken out. Thus the chlorophyll is almost
symmetric about the magnesium atom except
that there is a carbon
atom at one end and an oxygen atom at the
opposite end. For the
calculations performed on the analog structure,
the $Q_x$ and
$Q_y$ states are easy to identify.
However, for the calculations
based on the real chlorophyll structure,
where the electronic states are
mixed, this identification is difficult.
A fully symmetric
analog, with both oxygen atoms at the end,
was also proposed in
\cite{DAMJ2000A} based on the same
consideration and the error
arising from the simplified chlorophyll
analog was shown to be
insignificant compared to the
systematic errors.


\clearpage

\clearpage

\begin{figure}
\centerline{\includegraphics[width=16cm]{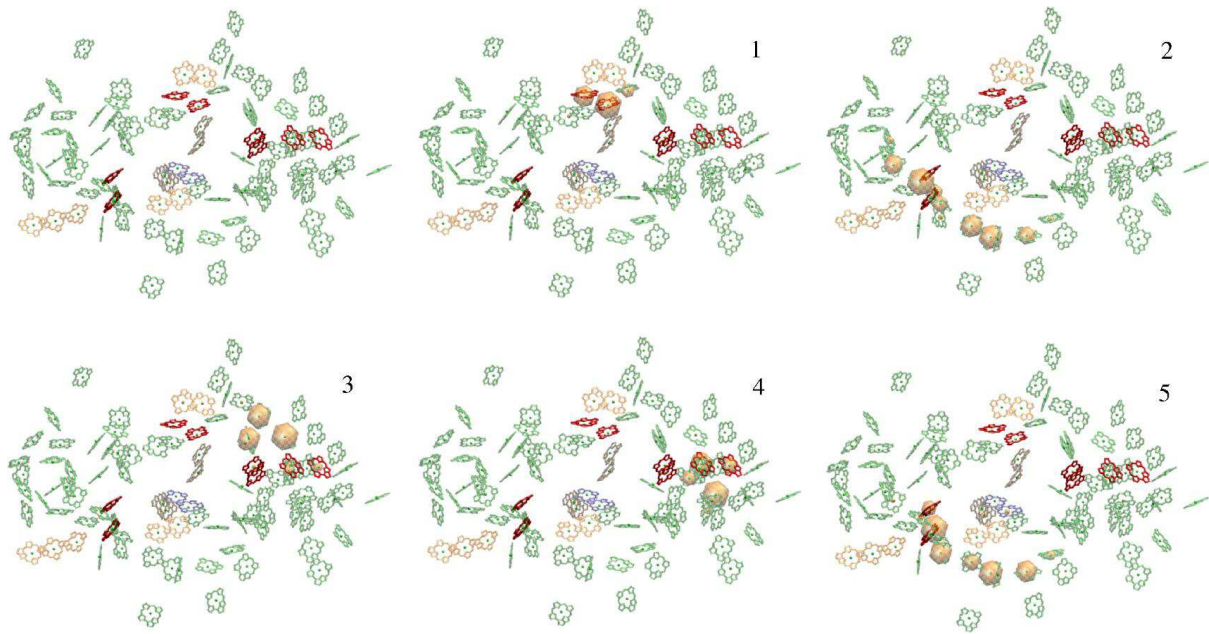}}
\caption[ps1fig]{ Excitonic (de)localization in PSI.
The first five eigenstates
of the effective Hamiltonian with full Coulomb couplings
and identical site energies are shown. The radius of the
sphere around a chlorophyll is proportional
to its occupation probability
in the respective state.
Excitons are delocalized
typically over no more than a few chlorophylls.
Figure produced with VMD~\cite{HUMP96}.}
\label{fig:ps1_eigs}
\end{figure}

\clearpage

\clearpage

\clearpage

\clearpage

\clearpage

\clearpage

\clearpage

\clearpage

\clearpage

\clearpage

\clearpage

\begin{table}
\caption{ Key chlorophyll
pairs in PSI and their couplings.  
Pairs
shown in boldface are the suggested red chlorophyll
candidates according to the
excitonic splittings computed in the
full Coulomb picture (see text for details).
FC: full Coulomb couplings,
DP: dipolar couplings, $O_{lo}/O_{hi}$: ratio of oscillator strengths
for the lower and higher excitonic states for full Coulomb couplings.
The couplings are given in units of~cm$^{-1}$. The pair
ecA1-ecB1 is the special pair P700.
Note that the dipolar approximation generally
overestimates the value of the coupling
at short distances.
The parentheses around the full Coulomb couplings
for the special pair and the trimer B31-B32-B33
indicate that these chlorophyll groups might have to be treated as
supermolecules, making the full Coulomb description
inaccurate 
(see text and Fig.~\ref{fig:trimer}).
}
\center{
\begin{tabular}{|c||c|c|c|} \hline
chlorophyll pair & $|H_{ij}|$ (FC) & $|H_{ij}|$ (DP) & $O_{lo}/O_{hi}$ \\
\hline
B37-B38 & 179 & 242 & 0.225  \\
A38-A39 & 162 & 199 & 0.171  \\
{\bf A32-B7} & {\bf 161} & {\bf 255} & {\bf 10.6} \\
{\bf A33-A34} & {\bf 153} & {\bf 193} & {\bf 14.3} \\
{\bf A24-A35} & {\bf 131} & {\bf 88.6} & {\bf 3.69} \\
{\bf B22-B34} & {\bf 127} & {\bf 92.6} & {\bf 3.50} \\
ecA1-ecB1 & (47.6) & 272 & 0.31 \\
B31-B32 & (88.8) & 301 & 191 \\
B32-B33 & (55.5) & 276 & 75.9 \\
\hline
\end{tabular}
}
\label{tbl:chl_pairs}
\end{table}

\clearpage

\begin{table}
\caption{  Average excitation lifetime and quantum yield.
{\em Model A:}  Full Coulomb couplings with P700 {\em tuned}
(see text) to the
observed absorption peak and with identical site energies (675~nm)
for the chlorophyll pool.
{\em Model B:} Dipole-dipole couplings with P700 tuned to the
observed absorption peak and with identical site energies (675~nm)
for the chlorophyll pool.
(The difference in lifetimes for models {\em A} and {\em B}
are mainly due to the difference of site
energy assignments for P700 for dipolar and full Coulomb
models.)
{\em Model C:} Full Coulomb couplings with P700 tuned to the
observed absorption peak, but with random site
energies (centered around 675~nm with a
width of 180~cm$^{-1}$) for the chlorophyll pool.
(Ensemble average value is given here.
For more detail see Fig.~\ref{fig:randomsite_hist}.)
{\em Model D:} Full Coulomb couplings with identical site
energies (675~nm) for all chlorophylls, including P700.
}
\center{
\begin{tabular}{|c||c|c|} \hline
{\em model} & {\em lifetime} (ps) & {\em quantum yield}  \\
\hline
{\bf A} & {\bf 27.4} & {\bf 97.3 \%  } \\
 B &  45.9 &  95.4 \%  \\
{\bf C} & {\bf 28.8} &  {\bf 97.1 \%} \\
D & 75.3 & 92.5 \%  \\
\hline
\end{tabular}
}
\label{tbl:tauandq}
\end{table}

\clearpage

\begin{table}
\caption{
 Characteristics of sojourn expansion.
The quantities given
(for definition see Fig.~\ref{fig:sojourn}
and Eq.~(\ref{deftausojsimple}))
have been evaluated
for the effective Hamiltonian
with full Coulomb couplings and adjusted P700
site energies (row 1, Table~\ref{tbl:tauandq}).
}
\center{
\begin{tabular}{|c|c|c|c|c|} \hline
 $\tau$ & $\tau_0$ & $\tau_{soj}$ & $q_{\bf 1}$
& $q_{T}$  \\ \hline
  27.6 ps &  11.8 ps
&  4.55 ps &   77.2~\%  &  77.8~\%  \\
 \hline
\end{tabular}
}
\label{tbl:sojourn}
\end{table}

\clearpage

\begin{table}
\caption{ Semiempirical parameters of
the PPP Hamiltonian as defined in expressions
 (\ref{eq_PPP1}), (\ref{eq_PPP2}), and (\ref{eq_PPP3}).}
\center{
\begin{tabular}{|rcl|rcl|rcl|}
\hline
\multicolumn{3}{|c|}{Carbon(C)}
&\multicolumn{3}{|c|}{Oxygen(O)}
&\multicolumn{3}{|c|}{Nitrogen(N)}
 \\ \hline
$I_k$&=&$11.16 \; {\rm eV}$ & $I_k$&=&$17.70 \;{\rm eV}$
& $I_k$&=&$14.12 \;{\rm eV}$ \\
$R_{kk}$&=&$11.13 \;{\rm eV}$ & $R_{kk}$&=&$15.23 \;{\rm eV}$
& $R_{kk}$&=&$12.34 \;{\rm eV}$ \\ \hline
\multicolumn{3}{|c}{}& \multicolumn{6}{l|}
{$\:\:\;\;\gamma_0\:\:=-2.43 \;{\rm eV}$} \\
\multicolumn{3}{|c}{}& \multicolumn{6}{l|}
{$\:\;\;Z_k\;\:=\;\;1.0$} \\
\multicolumn{3}{|c}{}& \multicolumn{3}{l}
{$\:r_{k,k \pm 1}\,=\;\;1.35 \; {\rm \AA}$}&
 \multicolumn{3}{c|}{(double bonds)}\\
\multicolumn{3}{|c}{}& \multicolumn{3}
{l}{$\;\;\:\;\;\;\;\;\;\:=\;\;1.46 \;
{\rm \AA}$}& \multicolumn{3}{c|}{(single bonds)} \\
\hline
\end{tabular}}
\vskip 4mm
\label{tbl:fcparams}
\end{table}


\end{document}